 \documentstyle[aps,preprint]{revtex}
\begin{document}

\title{  \bf \large 
Gap Opening Transition  and  Fractal Ground State Phase Diagram \\
in One Dimensional  
Fermions with Long Range Interaction : \\
 Mott Transition as a Quantum Phase Transition of Infinite Order }
\author{
Yasuhiro Hatsugai
}
\address{
 Department of Applied Physics,
 University of Tokyo 
 7-3-1 Hongo Bunkyo-ku, Tokyo 113, Japan
}
\maketitle
\begin{abstract}
 The metal-insulator transition in one dimensional fermionic systems 
with long-range  interaction is investigated.
 We have focused on an excitation spectrum by the exact diagonalization
technique
in   sectors with different momentum quantum numbers.
At  rational fillings, 
we have demonstrated  gap opening transitions 
from the Tomonaga-Luttinger 
liquid to the Mott insulator
associated with a discrete symmetry breaking
by changing the interaction
strength. 
  Finite interaction range is crucial
   to have the Mott transition at 
a rational filling away from the half filling.
 It is consistent with the strong coupling picture
 where the Mott gap exists at any rational fillings with sufficiently strong
 interaction. 
 The critical regions as a quantum phase transition are also investigated 
 numerically.
 Non-analytic behavior
 of the Mott gap 
is the characteristic in the weak coupling.
 It is of the order of
 the interaction in the strong coupling. 
It implies that the metal-insulator transition of the model is of the
infinite order as a quantum phase transition at
zero temperature. 
Fractal nature of the ground state phase diagram is also revealed.
\end{abstract}

\vskip 1.0cm

 \section{Introduction}

Effects of an electron-electron interaction in  electronic systems 
have become a focus of the condensed matter physics   recently.
In a three or higher dimensions, it is widely
 believed that   Landau's fermi liquid
theory is valid and the effect of the interaction is  
absorbed into several phenomenological Landau parameters.
The system is metallic even with the interaction since
the ground state is 
adiabatically connected to the free fermi sea where the excitations are
 given by an
electron-hole {\it gapless } excitation across the fermi surface. 
In one dimension, however, the ground state is unstable
against a perturbation of the interaction and 
the ground state is given by the so-called Tomonaga-Luttinger (TL) liquid.
Although the fermi liquid theory is not valid in one dimension, 
the TL liquid is also metallic and the excitations are gapless.
This TL liquid has been focused recently and there are huge number of studies
by several techniques as the bosonization  and the conformal field theory.
In the paper, we are trying to investigate the breakdown of the TL liquid 
behavior in
a simple fermionic system.
In a system with periodic potential, that is,
on a lattice as a model hamiltonian, 
the allowed kinetic energy is restricted by a finite
band width.
Therefore the 
 strong electron-electron
 interaction may  bring  an opening of the energy gap in  electronic systems
which is known   as a Mott transition.
 The opening of the energy gap implies
a metal-insulator transition which has become a focus again
recently in connection to new materials as metal-oxides, organic materials 
and the
high-$T_C$ superconductors.

As far as the conductivity of the electronic system 
 is concerned, the Mott transition can
be understood as a freezing of the charge degree of freedom. 
There can be spin related phenomena in a small energy scale, however,
the Mott transition is a phenomenon of the order of the electron-electron 
interaction.
 Noting this fact,
 we have focused on a charge degree of freedom and  chosen a model
hamiltonian of  spinless fermions on a lattice

\begin{equation}%
H=-t \sum_{j} c_{j+1}^\dagger  c_j +H.c. + \sum_{i\neq j} V_{| i-j |} n_i n_j,
\ (V_{k}\ge 0)
\label{eq:spls-hmlt}
\end{equation}%
where the interaction can be long range. ( We set $t=1$ in the following.)
When the interaction is non-zero only for
 the nearest-neighbor (NN) sites, it is mapped to the
spin
$1/2$  antiferromagnetic XXZ
model by the Jordan-Wigner transformation. 
In this case, Haldane investigated the model in detail by using the Bethe 
Ansatz
 solution of the XXZ model\cite{hldn-spls}.
Generic ground states 
of the NN model are the TL liquid without the energy gap
except at a half filling where the model has a metal-insulator transition at
$V_1=t$. 
It is identified as an antiferromagnetic  Ising gap in term of the XXZ spin
model. 

 In this paper, we  investigate the model, 
 when the interaction is   long range,
with various filling factors of the
fermion numbers.
As shown later, it has a rich structure as a {\em fractal}.

 \section{Strong Coupling}
Let us first consider a strong coupling limit $ V  >> t$ of the model when 
the filling factor $\rho=M/L$
is rational where $M$ is the number of   fermions and
$L$ is the number of   sites. 
We use a periodic boundary condition 
in the following.
Let us assume the interaction satisfies the downward convex condition,
that is, $(i+j)V_{l}\le j V_{l-i}+ i V_{l+j}$ ( $l<j,\ l,i,j <<L$ )
to avoid a formation of
charge clustering. (See later.) 
A possible form of the interaction  which we use in the paper is
\begin{eqnarray}%
V_{j} & = &   V f_j^L(\alpha) \\
f_j^L(\alpha) &= & \frac 1  {(\frac  L {\pi} \sin {\pi \frac j L})^\alpha},\
(\alpha\ge 1)
\label{eq:int-lng}
\end{eqnarray}%
which reduced to a simple power law decaying interaction 
$V_{j} = V/j^\alpha$ when $j<<L$ in a sufficiently large system. 
The nearest neighbor
interaction can be recovered also by taking $\alpha\to \infty$ and $L\to 
\infty$.
When the interaction is sufficiently large, the ground state of the system for
the rational filling $\rho= p/q$ with mutually prime integers $p$ and $q$ 
was known for
any $p$ and $q$.
It is given by a one dimensional Wigner crystal with period $q$\cite{hub}.
Although the ground state charge ordering ( crystal structure ) is 
complicated if $p$ and
$q$ are large integers,
it is explicitly given\cite{hub}.
 In   Fig.\ref{fig:chrg}, shape of the charge ordering is  shown
 for
 $\rho=p/q$ with $q=113$ and $p=1,\cdots,q-1$ as an example.
As is expected for the formation of the charge ordering, there are some 
commensurate
conditions to stabilize the system.
This commensurability condition  brings a fractal structure into the system
as shown later.
 The chemical potential in the thermodynamic limit
 is evaluated by a similar consideration
applied for the long range Ising model\cite{bak}. 
It is written as
\begin{eqnarray}%
\mu_\infty(\rho+0) & = & V  \sum_{k=1}^\infty e_+(k,\rho)  +F(\rho) \\
\mu_\infty(\rho-0) & = & V \sum_{k=1}^\infty e_-(k,\rho)  +F(\rho) \\
  e_+(k) & = & \left\{ \begin{array}{l}
 [\frac k \rho] f_{[\frac k\rho]-1}^\infty -  ([\frac k\rho]-1) 
f_{[\frac k\rho]}^\infty
{\rm \ if\ }  [\frac k\rho] = \frac k\rho 
\\
 ([\frac k\rho]+1)  f_{[\frac k\rho]}^\infty -  [\frac k\rho] 
f_{[\frac k\rho]+1} ^\infty
{\rm \ otherwise} 
\end{array}\right. \nonumber 
\\
 e_-(k) & = & 
 -[\frac k\rho] f_{[\frac k\rho]+1}^\infty +  ([\frac k\rho]+1) 
f_{[\frac k\rho]} ^\infty
\nonumber 
\end{eqnarray}%
where $[x]$ is a Gauss symbol which denotes a least integer which is not larger
that $x$ and $F(\rho)$ is an order
$t$ contribution mainly from the kinetic energy  which is a smooth function of
$\rho$.
In  Figs.~\ref{fig:chem}, the chemical potentials are evaluated for 
two different
interactions. 
It is a devil's staircase which was first discussed by Bak et al. in a context
of the long range Ising model\cite{bak}.
 The discontinuity of the chemical potential, $\Delta \mu  $,
 which is a key quantity to
judge whether the system is metallic ( $\Delta \mu = 0$ ) 
or not ($\Delta \mu >0$ )  is evaluated as
\begin{eqnarray}%
\Delta\mu_\infty
(\rho=\frac p q ) &= &  \mu_\infty(\rho+0) - \mu_\infty(\rho-0) 
=V g(\rho)\nonumber \\
 g(\rho=\frac p q) & = & \sum_{s=1}^\infty s q ( 
f_{sq+1}^\infty+f_{sq-1}^\infty-2 f_{sq}^\infty).
\label{eq:gap}
\end{eqnarray}%
This is generically non-negative for the potential which satisfies 
the convex downward condition. ( If this condition is not satisfied, 
$\Delta \mu$ can be negative for some filling factor which causes  
an instability.
It is identified as a charge clustering.) It implies that the energy gap opens  for
any rational fillings  if the interaction is strong enough. 
Further, as can be seen from Eq.~(\ref{eq:gap}), the energy 
gap $\Delta\mu(p/q)$ only
depends on $q$. 
Its dependence is given by a power law
$ \Delta\mu_\infty(\rho=\frac p q ) \sim \frac { Const. }{ q^\beta}
$.
From the consideration in the
   strong coupling limit, to have a Mott insulator phase in
the system with filling factor $\rho=p/q$, 
finite interaction range over $q$ sites is crucial. 

 \section{Weak Coupling}
On the other hand, when the interaction is weak ($V<<t$), 
the finite band width due to the lattice effect ( periodic potential )
is not so important.
In this case, one can approximate  the system as a continuous 
model with a long range 
 $g/r^\alpha$ interaction and with the periodic boundary 
condition\cite{suth2}.  
When $\alpha=2$, the continuum model is a Sutherland model 
which has been studied
extensively\cite{suth,hldn-an}. 
If $\alpha=2$, the weak coupling model can be discussed 
using an information from the exact
solutions.
 However,  the
strong coupling case, of course, neither the intermediate coupling 
case can not be 
discussed by the 
exact solutions.  
The ground state of the  Sutherland model is a TL liquid without an energy
gap independent of the filling
where the ground state is given by the Jastraw wavefunction. 
Noting that there is an energy gap in the strong coupling, 
it suggests that there is a
finite value of the interaction where the energy gap opens.
One possibility is that there is alway non-zero energy gap, that is,
the critical value where the energy gap opens, $V_C(\rho)=0$. 
However,  even in the nearest neighbor model where the energy gap can 
be most stable,
there is a gapless TL phase ( XY phase in the XXZ model ).
Therefore we can expect that 
there are always finite regions of the gapless TL liquid phases
in any filling factors and inevitably the  Mott transition.
In the following, we give  numerical results to confirm these consideration
 and study
the critical region.

 \section{Numerical results and Discussions}
As we have discussed, 
the ground state of
the spinless fermions  with long range interaction has two phases
for any rational filling when the interaction strength is varied. 
The one is the TL liquid metallic phase and the other is the
 Mott insulator phase.
The transition between the two 
 is a typical quantum phase transition at zero temperature.
In this section, these phase transitions are demonstrated numerically.
The main focus of the numerical calculations here is to investigate a
critical behavior near the gap opening ( the  transition point ).
Our main strategy is to investigate the system from an insulator side. 

We use the exact diagonalization technique for   systems with a periodic 
boundary
condition. The Hilbert space is divided into 
several sectors with different momentum  quantum numbers
and diagonalized within them
to obtain the lowest few   energies.
For  small systems, the full spectra   are also obtained. 
 Due to the small system size  available, it is not   efficient to calculate
 the
chemical potential directly. 
Instead of it, we calculate an excitation gap $E_{ex}$ which can be
comparable with    $\Delta \mu$. 

In Fig.~\ref{fig:disp1o2} and Fig.~\ref{fig:disp1o3},
lower parts of  energy spectra   are shown for systems with $\rho=1/2$ and
$\rho=1/3$, respectively. 
When the interaction is sufficiently weak, one  observes  a 
behavior  of  the gapless TL liquid as shown in  Fig.~\ref{fig:disp1o2}(a) and
Fig.~\ref{fig:disp1o3}(a). 
On the other hand,  opening of the energy gap near 
$k=  2 n k_F$, $n=0,\pm1, \pm2,\cdots$ ( $ = 2\pi m / q\ {\rm mod\ } q$,
 $m=1, 2,\cdots,q$
for
$\rho=p/q$) is clearly shown when
the interaction is sufficiently strong 
( See Fig.~\ref{fig:disp1o2}(b) and
Fig.~\ref{fig:disp1o3}(b)). 
As is known from the strong coupling analysis, 
there is a discrete symmetry breaking ( translational symmetry ) in 
the strong coupling
phase. For the $\rho=p/q$ case, this is a $Z_q$ symmetry breaking. 
Correspondingly there is almost $q$ degenerate ( ground ) states
in a finite system. 
They have different total momentum $2nk_F$ 
( $n=1,\cdots,q\ \ {\rm mod\ } q$ ) respectively.
At these momentum sectors, the lowest energy state is given by  
one of the $q$ degenerate ground states.
Therefore the lowest energy gap,
the energy difference between the lowest energy state at the momentum and  the
true ground state of the finite system ( usually $k=0$ ),
 is  related to an energy barrier between the degenerate $q$ ground states
barrier as shown in Eq.~(\ref{eq:gap-exp}). 
 The physical energy gap which we concern is the second lowest one as seen
 in the
Figures.
 In the Fig. ~\ref{fig:soft1o3}, the lowest energy gap at $k=2k_F$ for
$\rho=1/3$ is plotted as a function of the system size 
for several values of the interaction
strength.
 For $V=32$, the gap size obeys a exponential law as 
\begin{equation}%
E_{ex}(k=2k_F,L)  \sim e^{-c  L} 
\label{eq:gap-exp}
\end{equation}
which is a signature of the ( discrete ) symmetry breaking
and $c$ is  the order of the symmetry breaking potential.
The discrete symmetry breaking is confirmed numerically for $V=32$
in Fig.  ~\ref{fig:soft1o3}. 
To confirm the discrete symmetry breaking, 
we have also calculated a spectral flow, that is, 
the energy as a function of the Aharonov-Bohm flux
through the periodical system ( ring)\cite{kusakabe}.
In Fig.~\ref{fig:ab}, the spectral flows of 
three different momentum sectors ( $k=0,2\pi/3,$ and $4\pi/3$ ) for the
$\rho=1/3$  system is shown where the $Z_3$ symmetry breaking is expected. 
It is clearly shown that the $3$   low energy states are entangled with each
other. 
That is, these three states are equivalent in the thermodynamic limit which 
is a signature
of the discrete symmetry breaking.

Next let us investigate the critical region. 
For a nearest neighbor model at half filling,
 the behavior of the gap opening is known from
the Bethe Ansatz solution and is given by
 $\Delta\mu\sim \exp(-c/\sqrt{V-V_c})$ which is
essentially singular at the critical point\cite{yang-yang}.
We have numerically investigate the behavior of the gap opening 
for several filing factors in our long range interaction model.
In  Fig.~\ref{fig:gap1o2}, ~\ref{fig:gap1o3}, and  ~\ref{fig:gap1o4},
the excitation gap is plotted as a function of the interaction strength.

In the massless TL liquid phase 
and the Mott insulator phase with $ Z_q$ symmetry breaking, 
 $E_0(k=2\pi/L)-E_0(k=0)$ converges to the true energy gap 
in the thermodynamic limit. 
Therefore we extrapolate  $E_0(k=2\pi/L)-E_0(k=0)$, the energy gap
between the lowest energy at $k=0$ and
$k=2\pi/L$ to the infinite size limit by fitting them as a polynomial of the
$1/L$.

An example of the fitting is shown in Fig. ~\ref{fig:fit} for $\rho=1/3$ 
case with several
values of the interaction strength. 
The numerical results shown in Fig.~\ref{fig:gap1o2}, ~\ref{fig:gap1o3}, and 
~\ref{fig:gap1o4} can be well fitted by the
following essentially singular form 
as a function of the interaction strength 
\begin{equation}%
E_{ex}(V,\rho) \sim  c_0(\rho)\exp(-\frac {c(\rho)}
{(V-V_c(\rho))^{\gamma(\rho)}}), 
\ (V\ge V_c(\rho)).
\end{equation}%
It implies that 
 the transition is a generalized Kosterlitz-Thouless (KT) type. 
Therefore the transition is of   infinite order as a quantum phase transition.
This kind of the singular behavior is expected in the conformal field theory
for  the off-critical behavior near the conformally invariant critical phase.
In the $SU(\nu)$ invariant case, the exponent $\gamma$ is evaluated by the
renormalization group analysis to be $\gamma=\nu/(\nu+2)$\cite{itoi}. 
At the critical point of the model has an apparent $Z_q$ symmetry but may
 have higher
symmetry. Then one of the possible guesses for the exponent is
$\gamma(\rho=p/q)=q/(q+2)$. 

Numerically it is difficult to determine $V_c(\rho)$ and $\gamma(\rho)$ 
with sufficient
accuracies due to the singular behavior of the energy gap. 
However one may use 
$V_c(\rho)\sim t / g(\rho)$ as a lower bound of the $V_c$ since
it is plausible to expect the Wigner crystal at the strong coupling     melts 
near $t \sim \Delta\mu_\infty $. 
This estimate agrees  rather well with  numerical results obtained 
(Figs.~\ref{fig:gap1o2}, ~\ref{fig:gap1o3}, and 
~\ref{fig:gap1o4}).
In Figs. ~\ref{fig:phase}, we have shown possible ground state phase diagram
using the above estimate. 
In the figure, the interaction strength is plotted in a nonlinear scale. 
It shows that there is alway  insulator phase for any rational filling.
Since the $n-\mu_\infty $ curve is the
 devil's staircase, $\Delta \mu_\infty(\rho)$ shows a fractal structure.
Correspondingly, 
the ground state phase diagram of the spinless fermions with long range
 interaction
has a clear {\em fractal structure}\cite{spin}.

\section{  Acknowledgments}
The  author thanks Y. Morita,  M. Kohmoto, M. Takahshi, K. Maki and D. Lidsky 
 for helpful discussions.
This work was supported in part by Grant-in-Aid
from the Ministry of Education, Science and Culture 
of Japan.

 \begin{figure}
 \caption{
 Ground state charge ordering  of the one dimensional spinless fermions 
with long range
 interaction in the strong coupling. 
The black points are the positions of the particles. 
  $\rho=p/q$ with $q=113$ and $p=1,\cdots,q-1$.
The horizontal axis is the spacial direction $j$, $j=1,\cdots,q-1$. 
  \label{fig:chrg}}
 \end{figure}

\begin{figure}
\caption{
Evaluation of the chemical potential in the strong coupling limit. 
The smooth part $F(\rho)$ is set to be zero for simplicity.
 (a)
 $\alpha=2$,
 (b) $\alpha=1.5$.
 \label{fig:chem}}
\end{figure}

\begin{figure}
\caption{
Low energy spectra of the spinless fermions with long rang interaction  
classified by the momentum for  $\rho=1/2$.   
($\alpha=2$) 
The different symbols are for different system sizes; 
  $\triangle$: $L=10$,
  $\Diamond$: $L=14$,
  $\Box$: $L=18$,
 $\bigcirc$: $L=22$.
 (a) $V=1$. The thin lines are guide for  eyes. (b)   $V=4$. 
 \label{fig:disp1o2}}
\end{figure}

\begin{figure}
\caption{
Low energy spectrum of the spinless fermions with long rang interaction  
classified by the momentum for  $\rho=1/3$.   
($\alpha=2,\ t=1$) 
The different symbols are for different system sizes; 
  $\triangle$: $L=9$,
  $\Diamond$: $L=15$,
  $\Box$: $L=21$,
 $\bigcirc$: $L=27$.
 (a) $V=4$. The thin lines are guide for the eyes. (b)   $V=32$. 
 \label{fig:disp1o3}}
\end{figure}

\begin{figure}
\caption{
The energy gap at $k=2k_F$ is shown as a function of the system sizes for 
$\rho=1/3$ in log scale.    ($\alpha=2$) 
The different lines are for different values of the interaction strength;
$V=4$, $8$, $16$, and $32$ from above.
 \label{fig:soft1o3}}
\end{figure}

\begin{figure}
\caption{
The spectral flows ( energies as a function of the Aharonov-Bohm 
flux $\Phi$ through the
ring system)  of the three lowest energy states with momentum
$k=0,2\pi/3,$ and
$4\pi/3$ for the system with the filling factor
$\rho=1/3$ ($L=21$).    ($\alpha=2$ and $V=32$)
$\Phi_0$ is the flux quantum. When $\Phi=L\Phi_0=21\Phi_0$, 
the system returns to the original state by the (small) gauge invariance. 
 \label{fig:ab}}
\end{figure}

\begin{figure}
\caption{
The excitation gap  $E_{ex}$ as a function of $V$ for $\rho=1/2$. ($\alpha=2$) 
The data is taken by extrapolated to the infinite size.
 \label{fig:gap1o2}}
\end{figure}

\begin{figure}
\caption{
The excitation gap  $E_{ex}$ as a function of $V$ for $\rho=1/3$. ($\alpha=2$) 
The data is taken by extrapolated to the infinite size.
 \label{fig:gap1o3}}
\end{figure}

\begin{figure}
\caption{
The excitation gap  $E_{ex}$ as a function of $V$ for $\rho=1/4$. ($\alpha=2$) 
The data is taken by extrapolated to the infinite size limit.
 \label{fig:gap1o4}}
\end{figure}

\begin{figure}
\caption{
The excitation gap  $E_{ex}(L)$ as a function of $L$ for several values of $V$
( $\rho=1/3$). ($\alpha=2$)  
The solid lines are polynomial fits of the data in $1/L$.
 \label{fig:fit}}
\end{figure}

\begin{figure}
\caption{
Estimated ground state 
phase diagram of the spinless fermions with long rang interaction.
The solid lines are the  region with non-zero energy gap. ($\alpha=2$) 
The rest is a gapless TL liquid.
The scale in the vertical axis is non-linear as $\frac {\xi^V-1}{\xi^V+1}$;
(a) $\xi=1.15$ and (b) $\xi=1.01$ ( The strong coupling region is expanded).
 \label{fig:phase}}
\end{figure}


\begin{references}
\bibitem{spin}Spin model with inverse square exchange with anisotropy
also has a similar fractal structure\cite{bak} 
by the similar argument given in the
manuscript.  
Since {\em only a  part} of the eigenstates is obtained analytically\cite{hldn-spls},
the fractal nature of the system is not captured
 by the analytical solutions\cite{hah,saiga}.
\end{references}
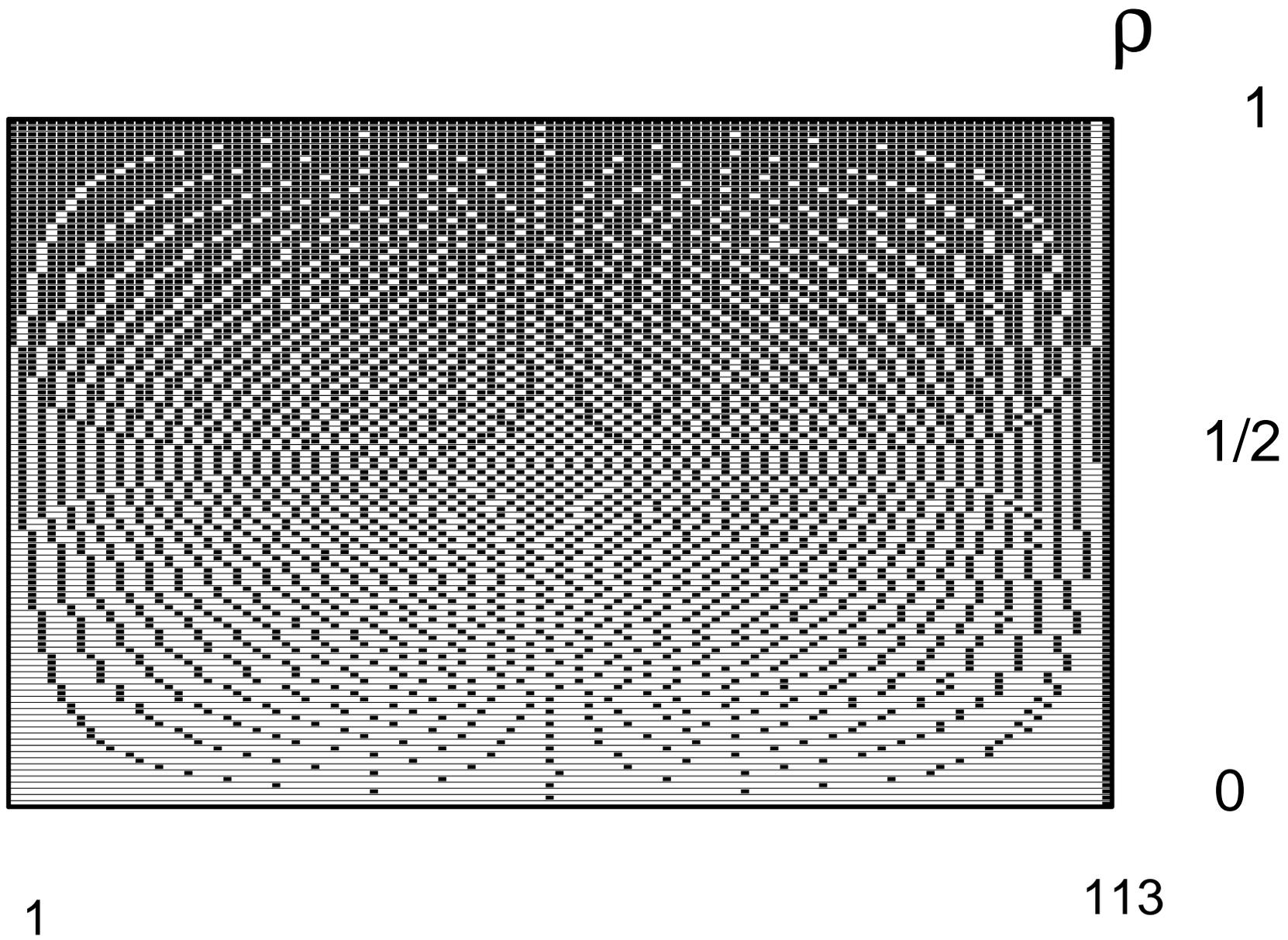
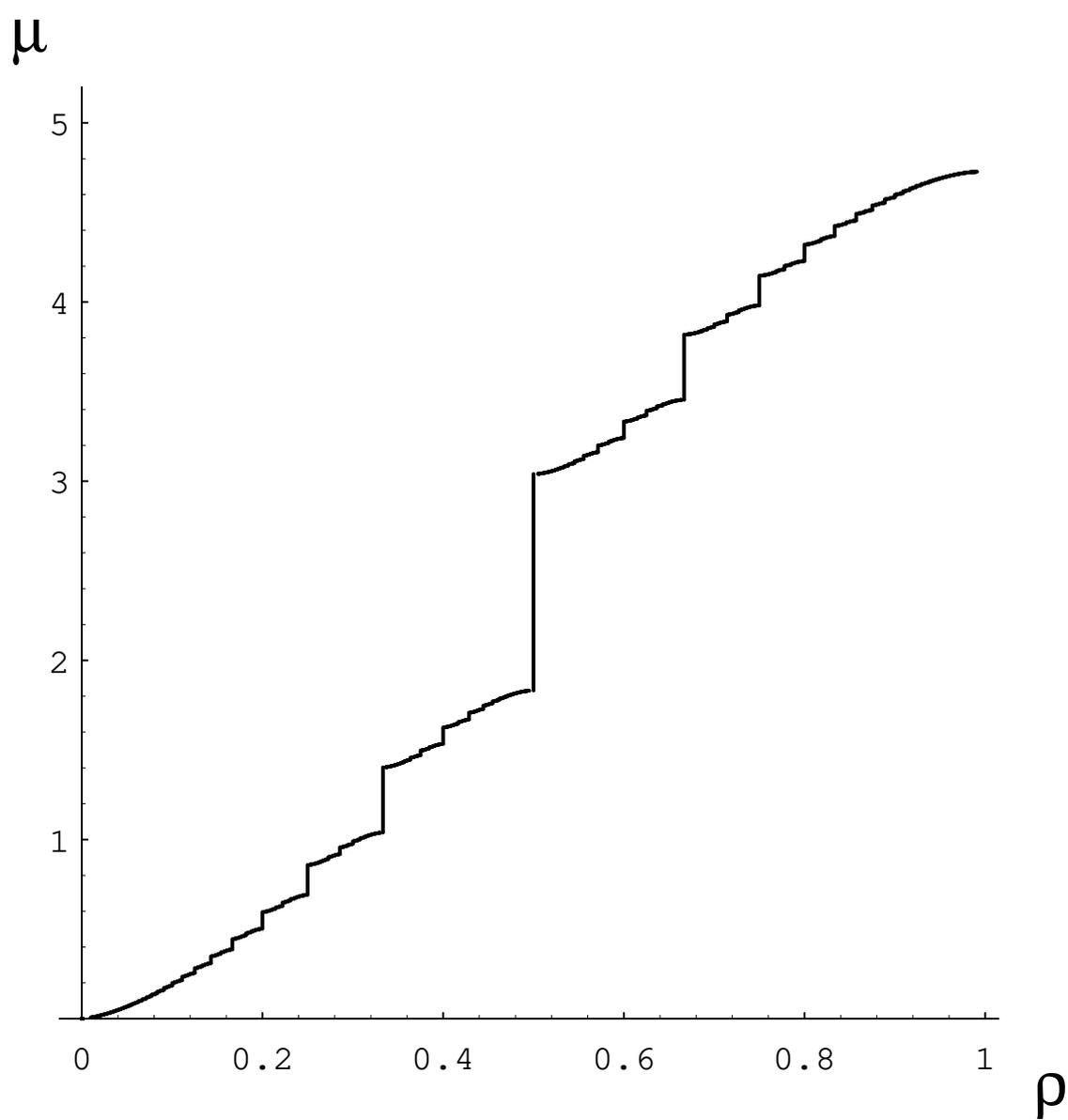
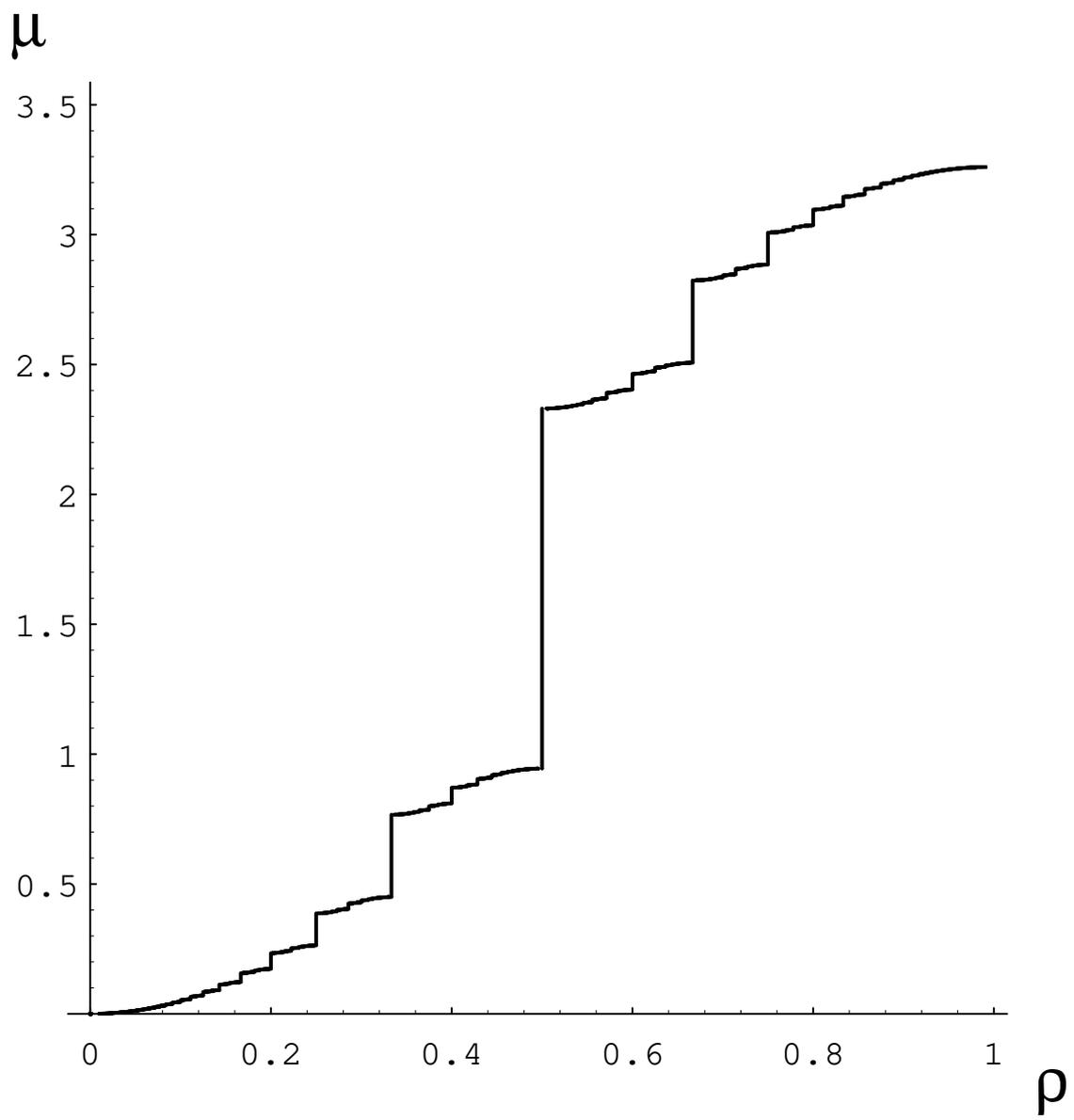
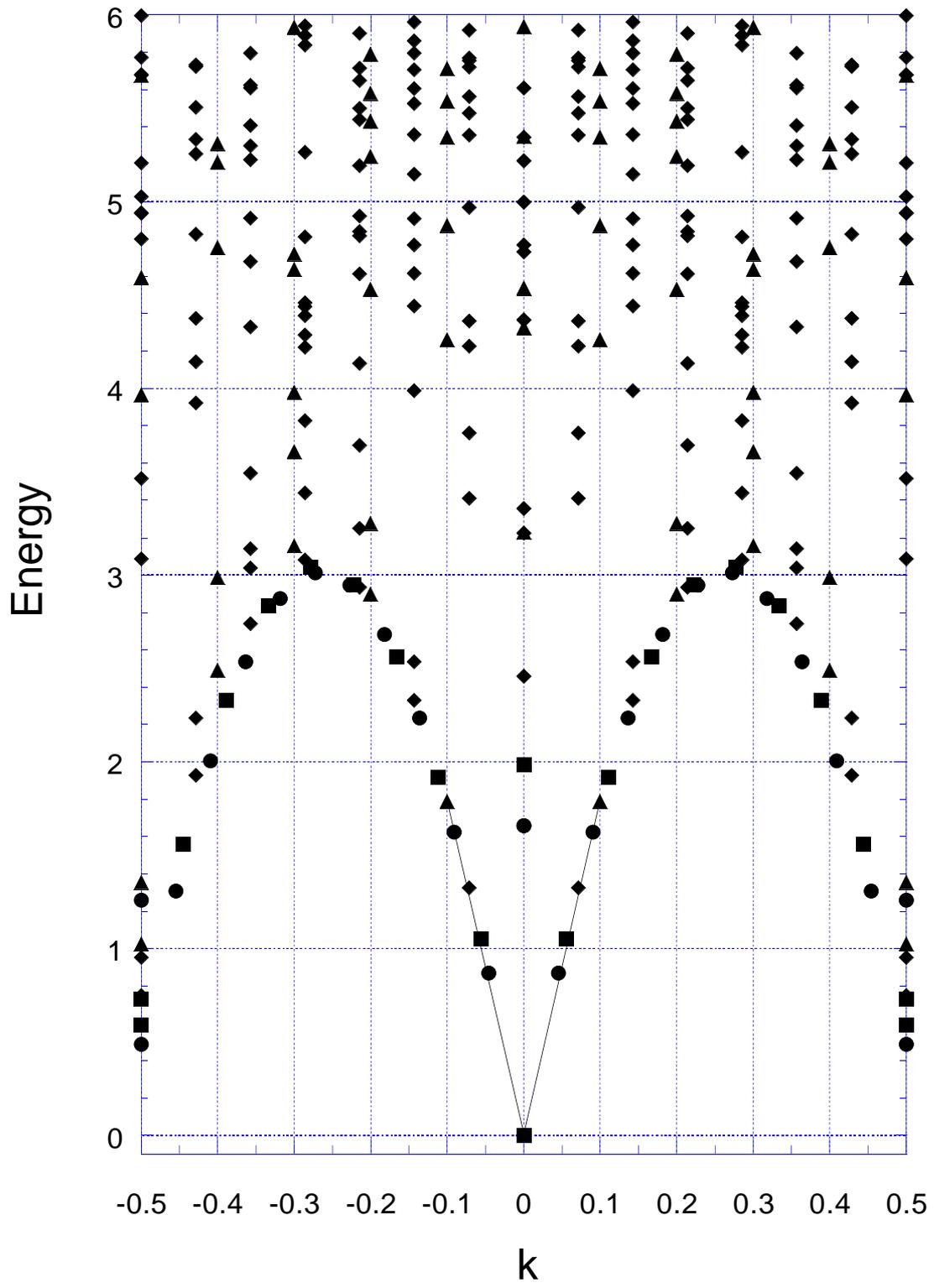
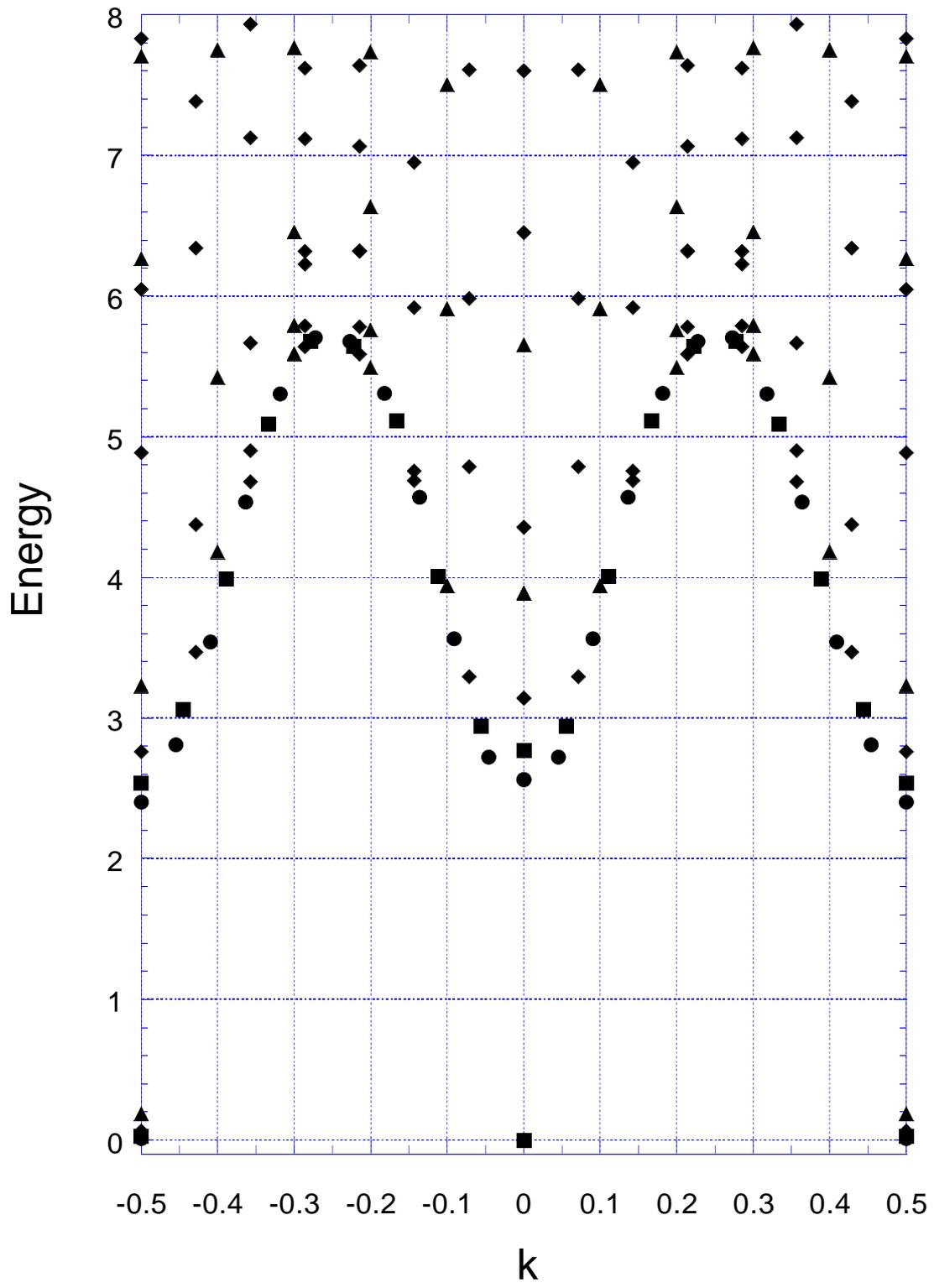
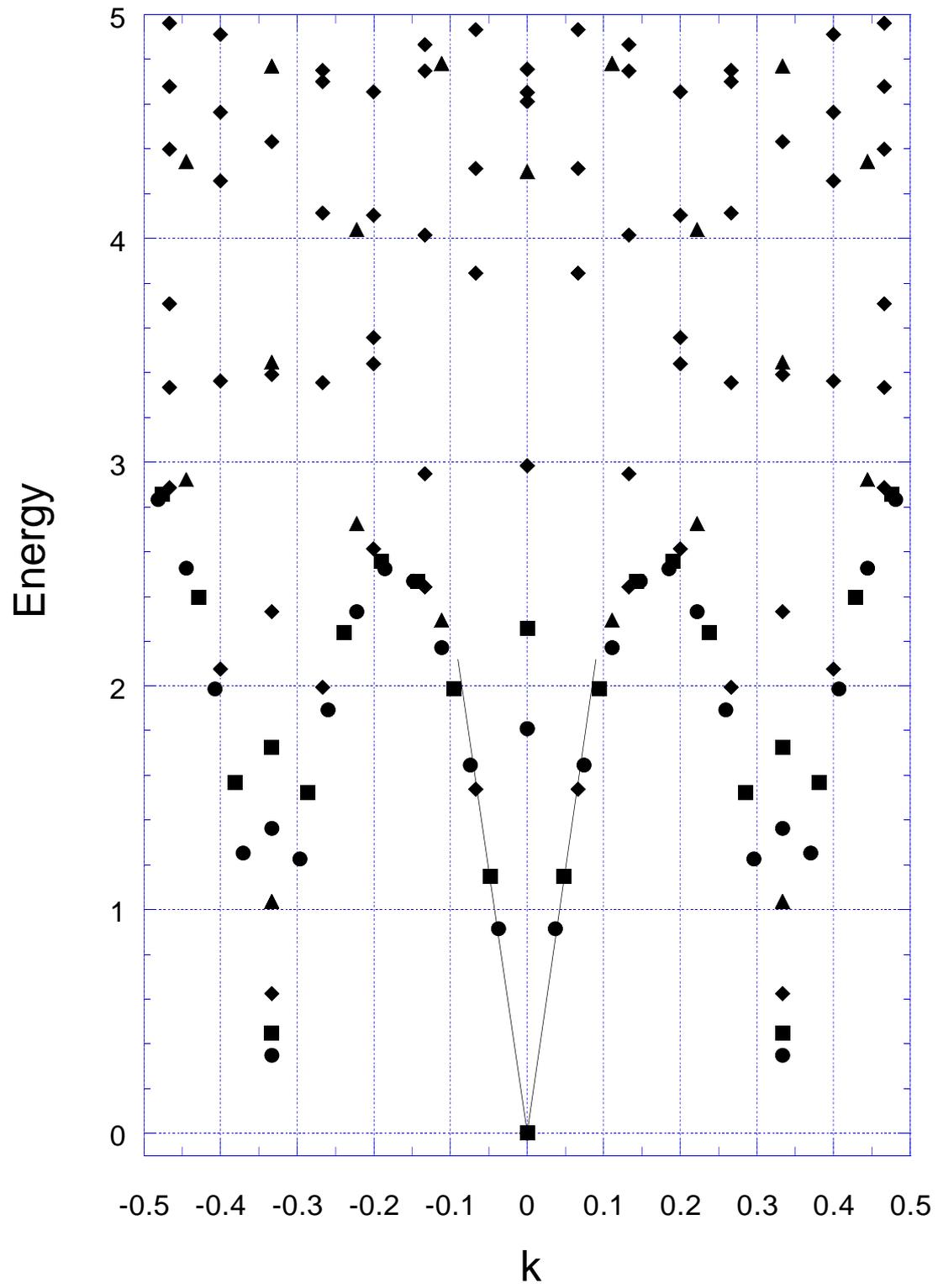
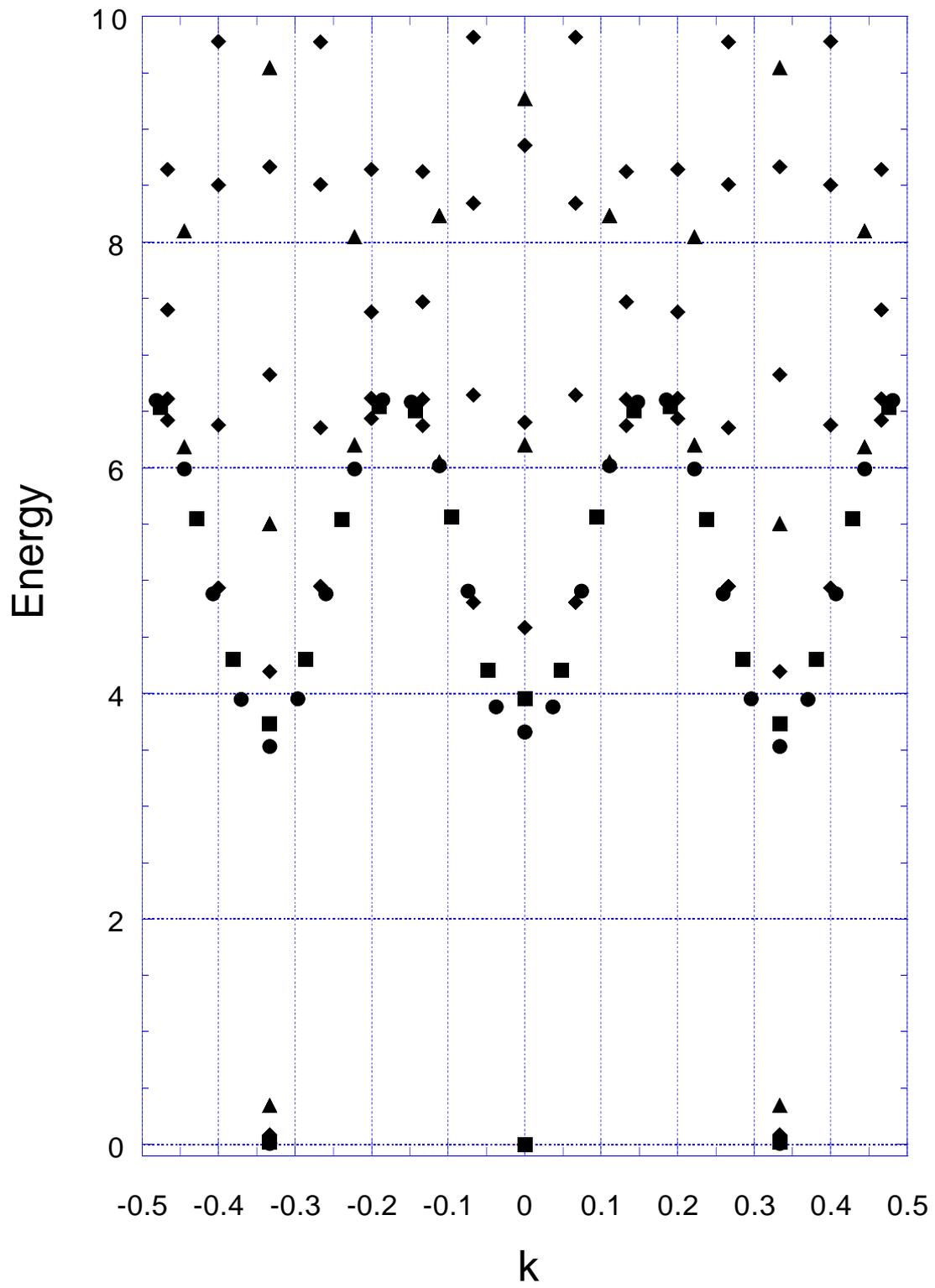
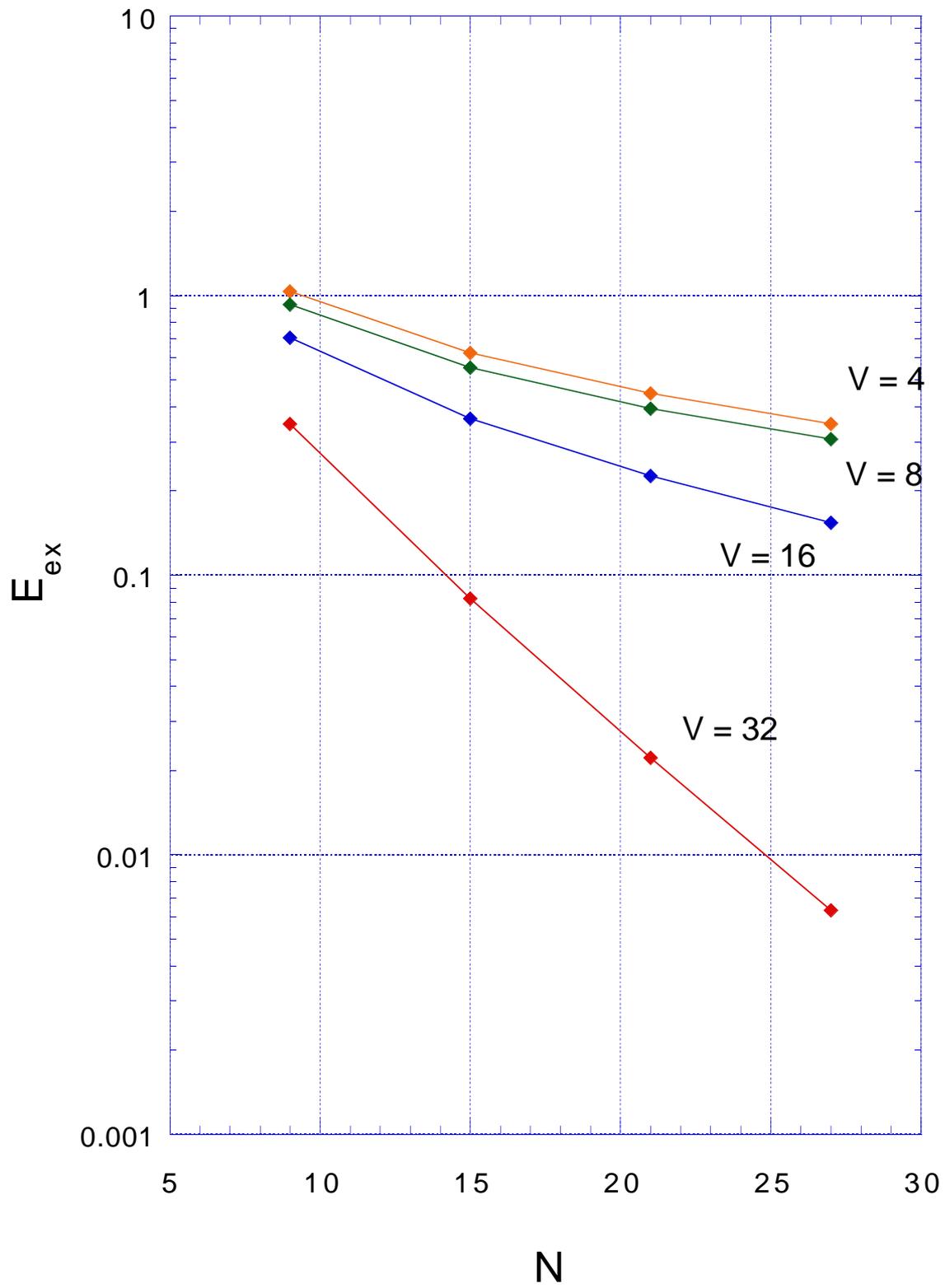
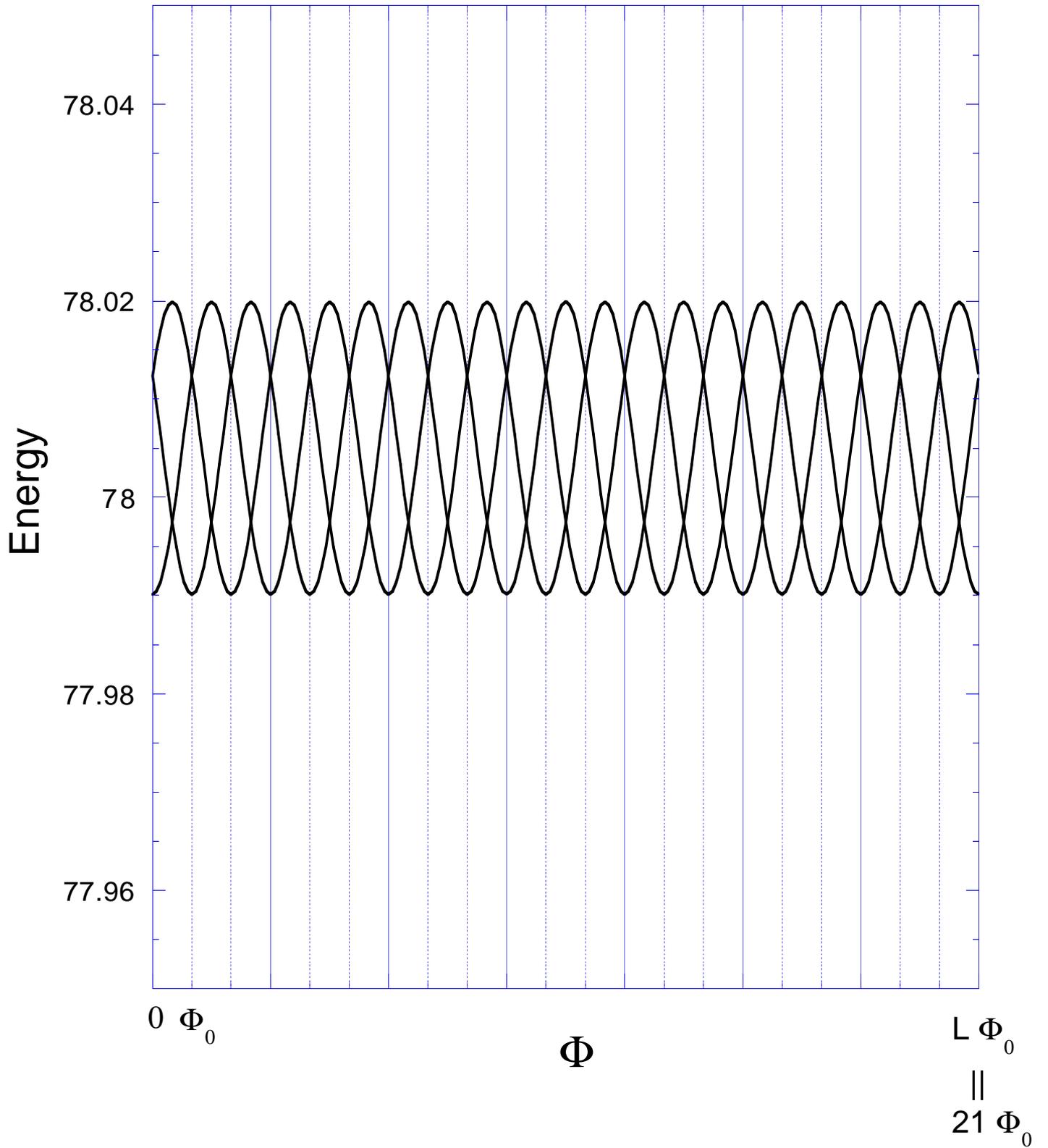
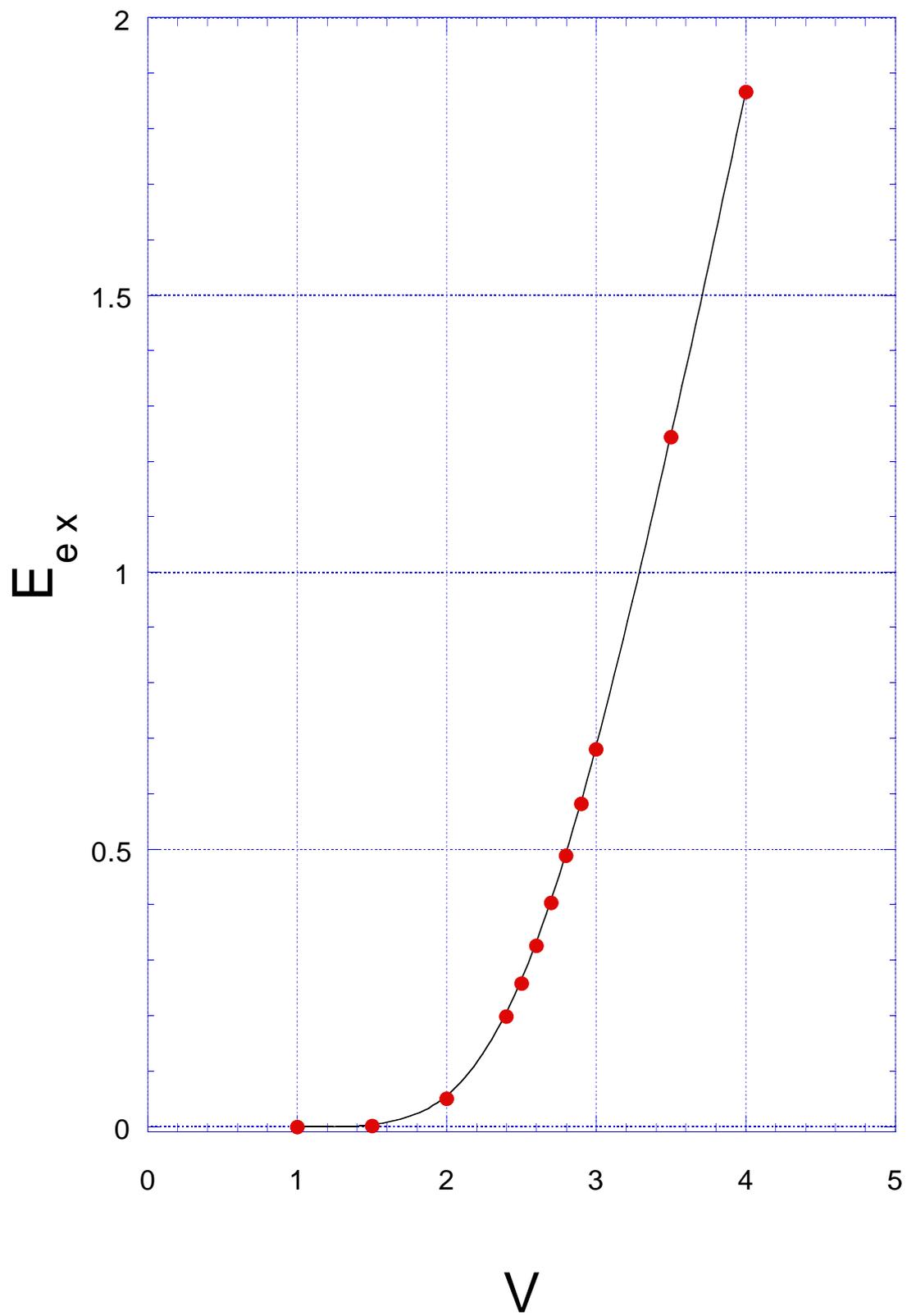
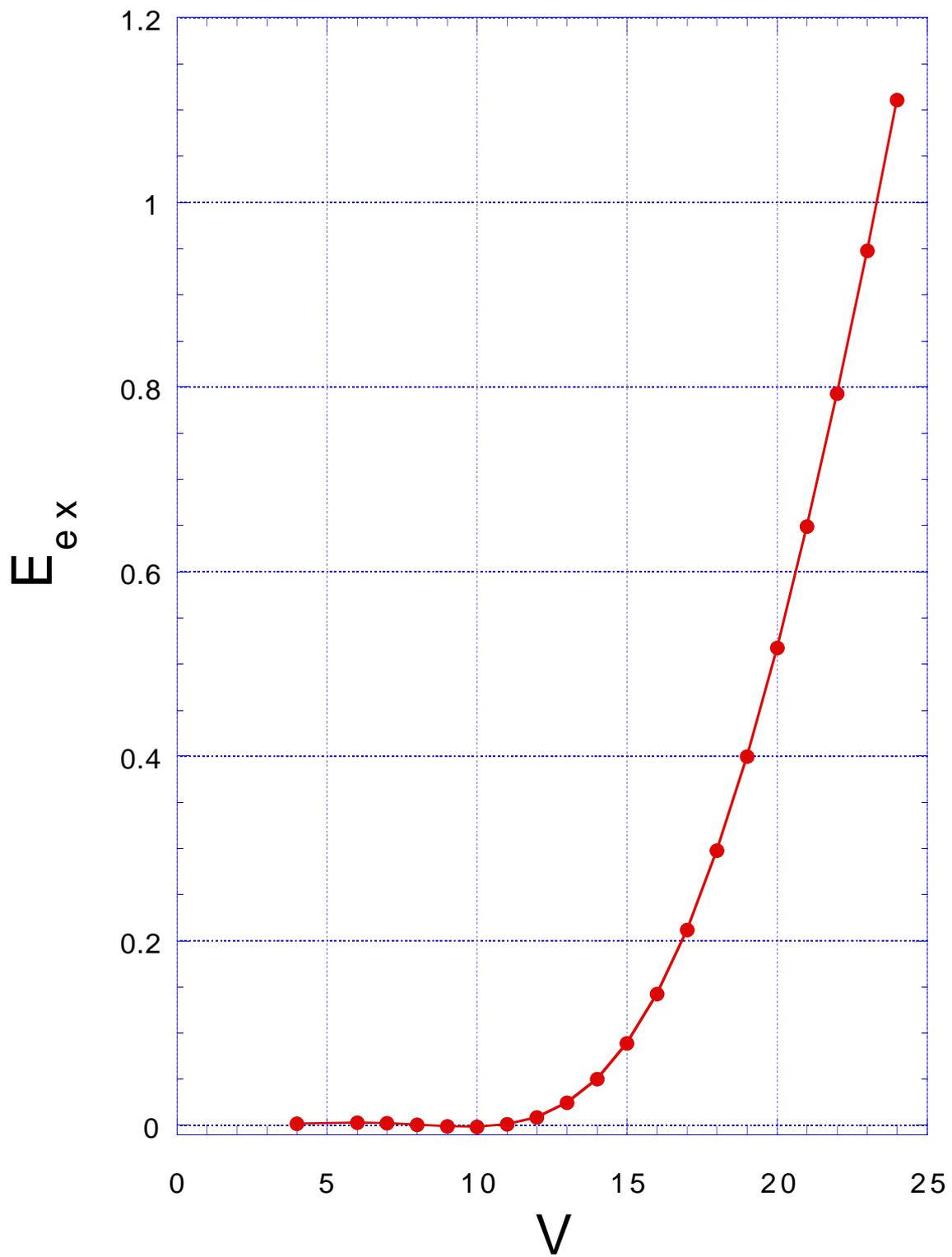
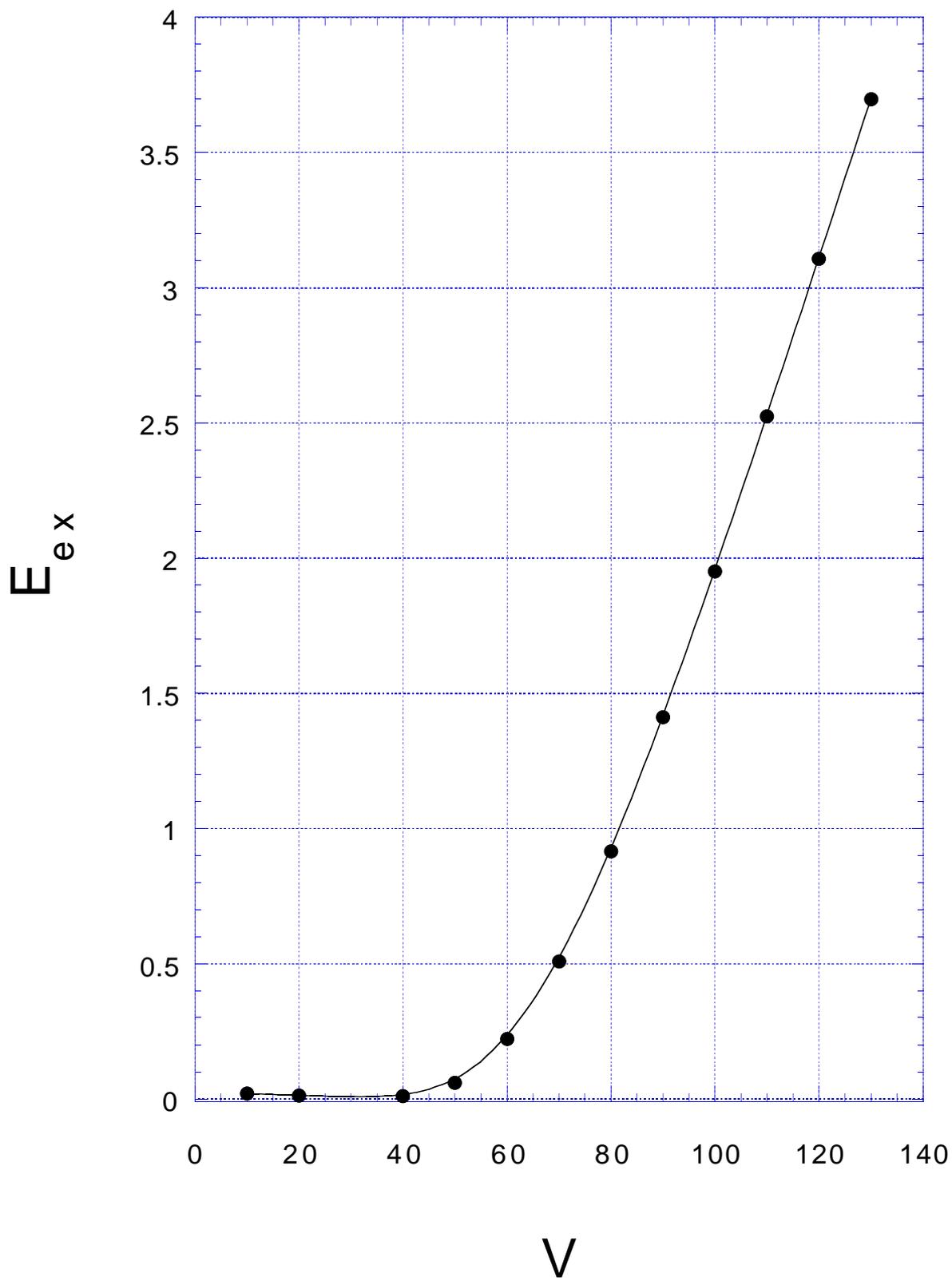
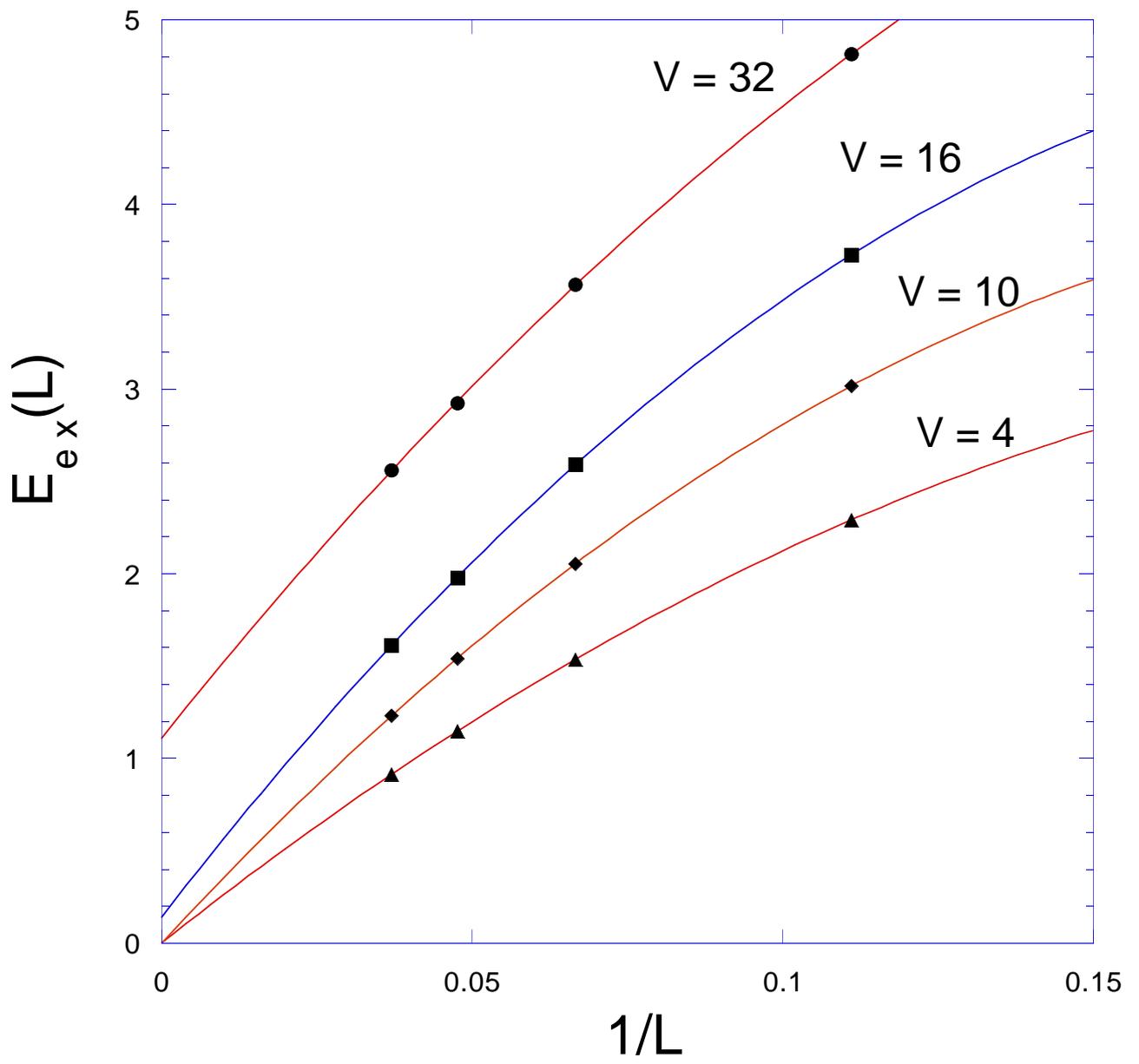
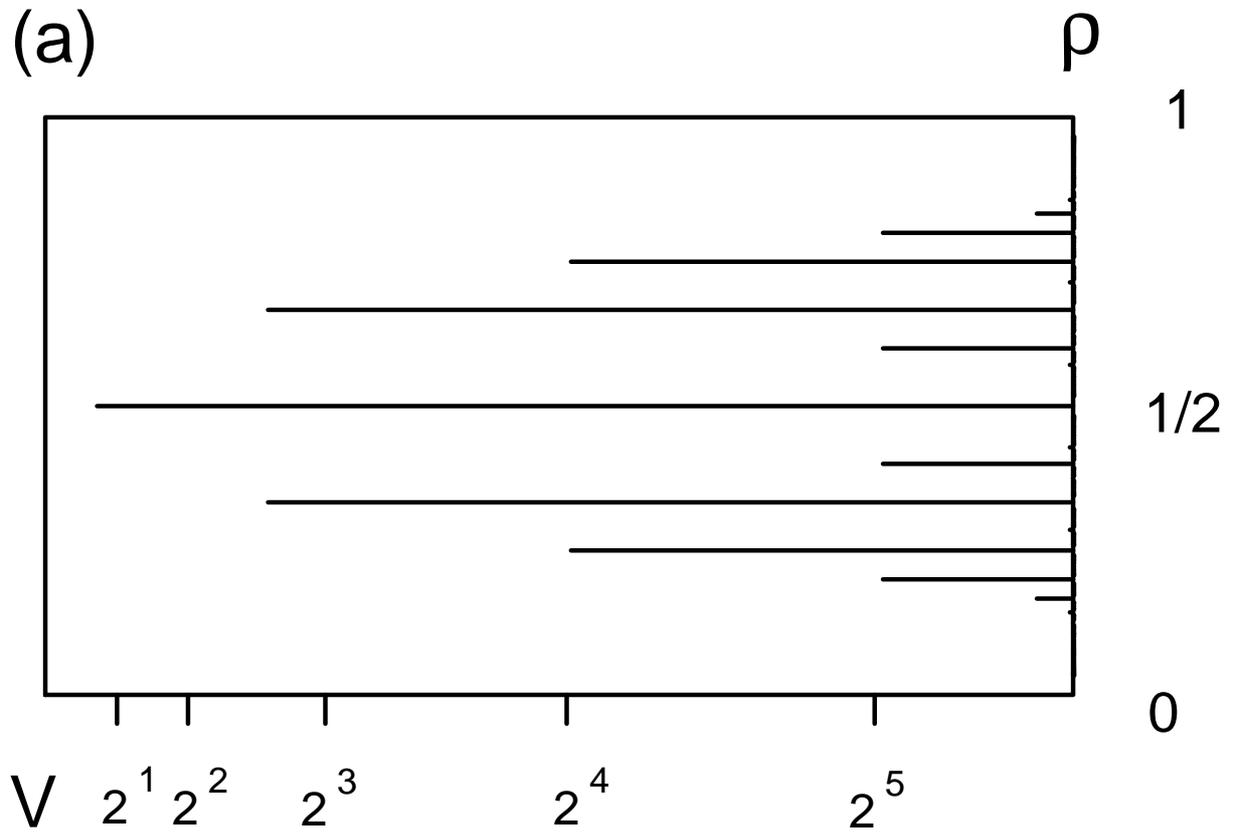
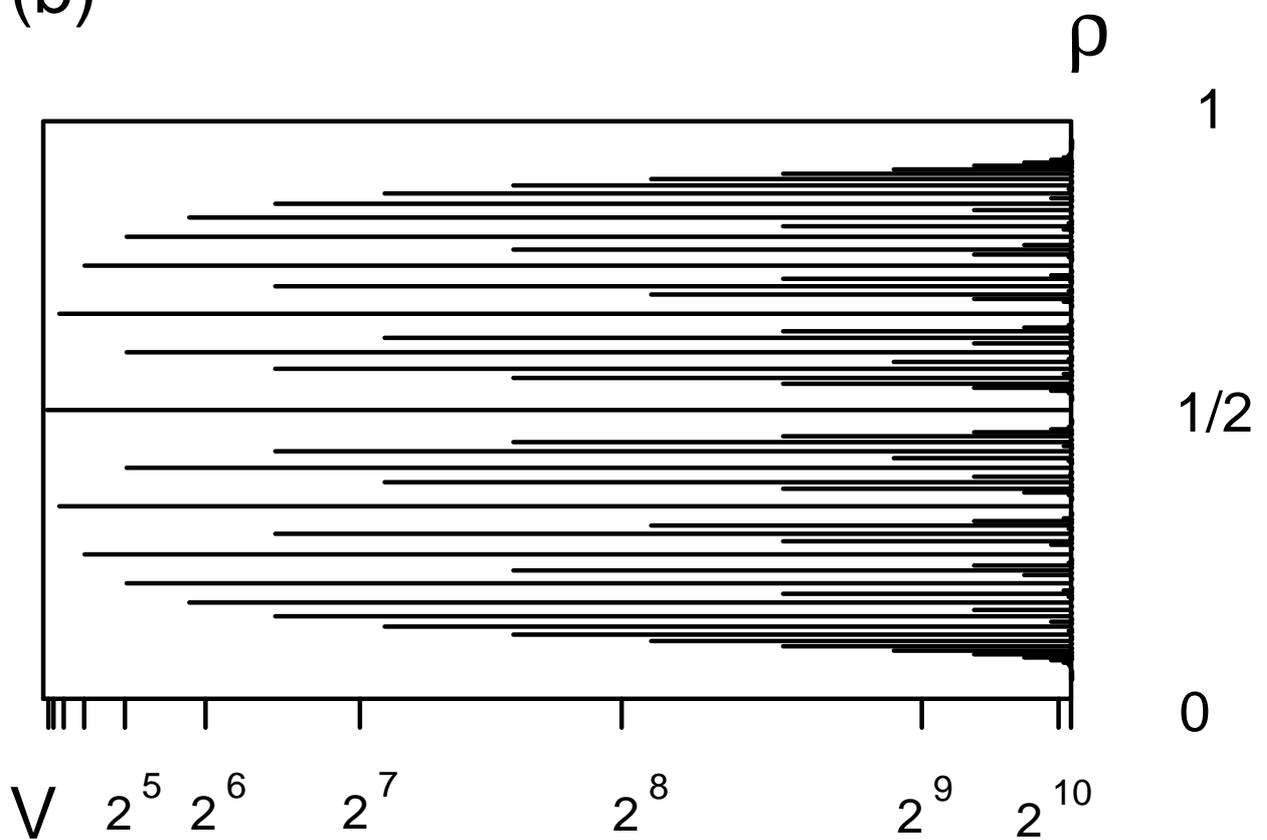
\end{document}